\documentclass[aps,pre,preprint,showpacs,showkeys]{revtex4-1}

\usepackage{setspace}
\usepackage{amstext}
\usepackage{amsmath}%
\usepackage{amsfonts}%
\usepackage{amssymb}%
\usepackage{amsbsy}
\usepackage{amsgen}
\usepackage{color}

\usepackage{graphicx}

\newlength{\fw}
\begin{document}
\title{Transport of Brownian particles in a narrow, slowly-varying serpentine channel}

\author{Xinli Wang}
\affiliation{Division of Mathematics and Computer Science\\
University of South Carolina Upstate}

\author{German Drazer}
\affiliation{Mechanical and Aerospace Engineering\\ Rutgers, the State University of New Jersey}

\date{\today}

\begin{abstract}
We study the transport of Brownian particles under a constant driving force and moving in channels that present a varying centerline but have constant aperture width.
We investigate two types of channels, {\em solid}
channels in which the particles are geometrically confined between walls and {\em soft} channels in which
the particles are confined by a periodic potential. We consider the limit of narrow, slowly-varying channels,
i.e., when the aperture and the variation in the position of the centerline
are small compared to the length of a unit cell in the channel (wavelength).
We use the method of asymptotic expansions to determine both the average velocity (or mobility) and the effective diffusion coefficient
of the particles.
We show that both solid and soft-channels have the same effects on the transport properties up to $O(\epsilon^2)$.
We also show that the mobility in a solid-channel at $O(\epsilon^4)$ is smaller than that in a soft-channel.
Interestingly, in both cases, the corrections to the mobility of the particles are independent of the P\'eclet number and,
as a result, the Einstein-Smoluchowski relation is satisfied.
Finally, we show that by increasing the solid-channel width from $w(x)$ to $\sqrt{6/\pi}w(x)$, the mobility of the particles in
the solid-channel can be matched to that in the soft-channel up to $O(\epsilon^4)$.
 \end{abstract}


\maketitle

\section{Introduction}
The diffusive transport of suspended particles confined to channels is important in a wide range of problems that take place
both in natural systems, e.g., particle transport in cells \cite{AlbertsJLRRW2007} and modeling drug delivery \cite{Saltzman2001},
as well as in engineered systems, such as in the development of separation and analytical microfluidic devices  \cite{Duke1998,Pamme2007,HerrmannKD2009,2010-Koplik-PoF,BernateD2011,2012-Jorge-PRL,bernate_vector_2013}.

The unbiased Brownian motion of suspended particles confined to a symmetric channel (or pipeline) with hard walls has been extensively studied,
using for example the Fick-Jacobs (F-J) approximation \cite{Jacobs1967}, which reduces the dimensionality of the problem by averaging over the cross
section.
Zwanzig later modified the F-J approximation with a position-dependent effective diffusion coefficient that takes into account the curvature of
the confining boundary \cite{Zwanzig1992}.
Reguera and Rubi proposed a scaling law for the effective diffusion coefficient in order to improve
the approximation in the case of boundaries with significant curvature (important variations in the aperture of the channel) \cite{RegueraR2001,BuradaSRRH2007}.
Using a different approach, based on a projection method, Kalinay and Percus systematically derived the projected
one-dimensional problem by assuming that the diffusion time in the transverse direction is much smaller than that in the longitudinal
direction \cite{KalinayP2005,KalinayP2006}. This approach is equivalent to an asymptotic expansion on the width of the channel and, in principle,
it can be used to derive all higher order corrections to the F-J approximation. Note that in this case the centerline of the channel is a straight line and
the corrections result from the variations of the aperture of the channel.
Also important, particularly in the context of microfluidic devices, is the diffusion of suspended particles in a channel of constant width but with a position of the varying centerline
(sometimes called a serpentine channel \cite{RushDBK2002}). In this case, corrections based on the derivatives of the channel aperture would clearly vanish.
Bradley \cite{Bradley2009} performed an asymptotic expansion on the width of channel and derived an expression for the effective diffusivity of particles
confined to a narrow channel, both in the case of a serpentine channel as well as for a channel with a varying aperture. More recently, Dagdug and Pineda showed that the same results can be obtained by the projection method \cite{DagdugP2012}.
This approach was also generalized to an arbitrary multidimensional system by Berezhkovskii and Szabo \cite{BerezhkovskiiS2011}.

In recent years, the biased transport of suspended particles in the presence of an external force field has received considerable attention due to the development of novel
separation strategies in microfluidic devices. In the simplest case, in which the external force is constant in the longitudinal direction, a straightforward
extension of the F-J approximation has been used to evaluate the average velocity and the effective diffusivity of Brownian
particles  \cite{BuradaHMST2009,RegueraSBRRH2006, BuradaSRRH2007}.
Alternatively, asymptotic analysis has also been used to calculate effective (or {\it macro-}) transport properties of Brownian particles confined either to
a narrow channel \cite{LaachiKYD2007} or to a weakly corrugated one \cite{WangD2010}.
Analogous behavior is observed in the biased transport of Brownian particles confined by
a potential energy landscape (a {\it soft} channel) \cite{WangD2010}, which is relevant
to partition-induced separation in microfluidic devices \cite{DorfmanB2001, BernateD2011}.
In the case of soft-channels, we have previously shown that, the leading order effect of the confining potential on the transport properties of the
suspended particles is the same as that induced by solid walls, as long as the entropic
barriers created by the varying aperture of the channel are the same \cite{LaachiKYD2007,WangD2009,WangD2010}.

Here we extend previous work to consider the case of biased transport of Brownian particles in a channel of constant width but varying centerline.
In particular, we use  asymptotic analysis to investigate the leading order correction to the effective diffusion coefficient. We also calculate higher order terms in the asymptotic expansion of the average velocity (or mobility). We consider both a solid-channel as well as a channel created by a confining potential, and compare the results.

\section{Transport of Brownian particles in a curved channel}

Let us first describe the geometry of the channels considered in this work. There are three important characteristic length scales: $L$ -- the length of one period in the longitudinal direction, $a$ -- the average channel width, and $\delta_z$ -- the amplitude of the variation in the position of the boundaries. Then, the problem can be categorized into three main cases:
a {\it slowly-varying} channel for $\delta_z/L\ll1$,
a {\it narrow} channel for $a/L\ll 1$,
and a {\it weakly corrugated} channel for $\delta_z/a\ll 1$.

Here we study the biassed motion of a Brownian particle in a {\it narrow, slowly-varying} channel ($a/L\ll1$ and $\delta_z/L \ll1$, $\delta_z/a \sim O(1)$) with
a constant aperture but varying centerline. The bias is induced by a constant and uniform external force acting in the longitudinal direction and
we consider both soft and solid-channels.
In the case of a soft-channel, Brownian particles are confined by a potential that is  periodic in the $X$-direction, $\bar {V}(X,Z)=\bar V(X+L,Z)$,
and confines the particles in the $Z$-direction, $\bar V(X,Z)\to+\infty$ for $Z\to \pm \infty$.
On the other hand, in the case of a solid-channel, Brownian particles are confined between solid walls, described by $Z=Z_+(X)$ and $Z=Z_-(X)$.
(Note that the potential $\bar V(X,Z)=0$ when considering the transport in a solid-channel.)

In the limit of negligible inertia effects, the motion of the particles is described by the Smoluchowski equation for the probability density $\bar P(X,Z,t)$,
\begin{equation}
\frac{\partial \bar P}{\partial t}+\nabla\cdot {\bf \bar J}=\delta(X,Z)\delta(t).
\end{equation}
The probability flux, ${\bf \bar J}(X,Z,t)$, is given by
\begin{equation}
\label{J} {\bf \bar J} = \frac{1}{\eta}\left[\left(F -
\frac{\partial \bar V}{\partial X}\right) \bar P -
k_BT\frac{\partial \bar P}{\partial X}\right]\vec{i} +
\frac{1}{\eta}\left(-\frac{\partial \bar V}{\partial Z}\bar
P-k_{B}T\frac{\partial \bar P}{\partial Z}\right)\vec{k},
\end{equation}
where $\vec{i}$ and $\vec{k}$ are the unit vectors along $X$ and $Z$, respectively, $\eta$ is the viscous
friction coefficient, $F$ is a uniform external force in the $X$-direction, $k_B$ is the Boltzmann constant, $T$ is the absolute temperature,
and the Stokes-Einstein equation is used to write the diffusion coefficient in terms of $\eta$, $D=k_{B}T/\eta$.
 The inertia effects are negligible and, therefore, the velocity is simply the ratio of the force to the viscous friction coefficient $\eta$.

Instead of considering the problem in an unbounded domain in $X$, it is convenient to
introduce the reduced probability density (and probability flux) which maps the infinite domain into
a single period of the channel (see Refs. \onlinecite{Reimann2002, LiD2007}),
\begin{eqnarray}
\label{reduced}
\tilde P(\bar x,\bar z,t)= \sum_{n_x=-\infty}^{+\infty} \bar P(\bar x+n_x L,\bar z,t), \\
\mathbf{\tilde{J}}(\bar x, \bar z, t)= \sum_{n_x=-\infty}^{+\infty}\mathbf{\bar J}(\bar x+n_xL,\bar z,t),
\end{eqnarray}
where the integer $n_x$ indicates the number of periods along the channel, and $(\bar x, \bar z)$ is the same to $(X, Z)$ except that $\bar x$ is defined in $[0,L]$. The reduced probability density can then be obtained
by solving the Smoluchowski equation with periodic boundary conditions in $\bar x$. In particular, the long-time asymptotic
probability density, $\tilde P_{\infty}(\bar x,\bar z)=\lim
_{t\to\infty}\tilde P(\bar x,\bar z,t)$, is governed by the
equation
\begin{equation}
\nabla\cdot {\bf \tilde J}_{\infty} = 0,
\label{eqn:governing_original}
\end{equation}
with the normalization condition for the reduced probability density,
\begin{equation}
\langle\tilde P_{\infty}\rangle\stackrel{\text{def}}{=}\int\int_{\Omega}\tilde P_{\infty} \mathrm{d}\bar
x\mathrm{d}\bar z = 1,
\label{eqn:normalization}
\end{equation}
where $\Omega=\{(\bar{x}, \bar{z}): 0\leq\bar{x}\leq L, -\infty<\bar{z}<\infty \}$ for a soft-channel and $\Omega=\{(\bar{x}, \bar{z}):0\leq\bar{x}\leq L, \bar{z}_-\leq\bar{z}\leq\bar{z}_+\}$ for a solid-channel.
The boundary conditions are periodic in $\bar x$,
\begin{equation}
\tilde P_{\infty}(0,\bar z)=\tilde P_{\infty}(L,\bar z),
\label{eqn:period}
\end{equation}
and the zero-flux condition in $\bar{z}$, which depends on the type of the channel.
For a soft-channel, it corresponds to a vanishingly small probability density and flux in the limit of large $\bar{z}$ values,
\begin{equation}
\tilde J_{\infty}^{\bar z} = \frac{1}{\eta}\left(-\frac{\partial
\bar V}{\partial \bar z}\tilde P_{\infty}-k_{B}T\frac{\partial
\tilde P_{\infty}}{\partial \bar z}\right) \xrightarrow[{\bar z
\to \pm \infty}]{}  0. \label{eqn:flux}
\end{equation}
In the case of a solid-channel, the zero-flux condition at the boundaries $\bar z=\bar z_{\pm}$ is given by
\begin{equation}
\tilde{J}_{\infty}\cdot\vec{N}=0,
\end{equation}
where $\vec{N}$ is the vector normal to the channel walls.

Let us now introduce the following dimensionless variables using the characteristic scales of the problem,
$x=\bar x/L$, $z =\bar z/a$, $V = \bar V/(k_B T)$, as well as the re-scaled probability density $P_{\infty}=a L\tilde P_{\infty}$.
The governing equation for the reduced probability then becomes:
\begin{equation}
\epsilon^{2}\frac{\partial}{\partial
x}\left[\left(\textrm{Pe}-\frac{\partial V}{\partial x}\right)
P_{\infty}-\frac{\partial P_{\infty}}{\partial x}\right] +
\frac{\partial}{\partial z}\left[-\frac{\partial V}{\partial
z}P_{\infty}-\frac{\partial P_{\infty}}{\partial z}\right] = 0,
\label{eqn:governing}
\end{equation}
where $\epsilon =a/L$ is the aspect ratio of the channel, and $\textrm{Pe}=FL/k_B T$ is the P\'eclet number (a measure of the relative  importance of convective and diffusive transport). The boundary conditions in dimensionless form are: the periodic boundary condition in the $x$ direction,
\begin{equation}
P_{\infty}(0,z)=P_{\infty}(1,z),
\end{equation}
the normalization condition,
\begin{equation}
\langle P_{\infty}\rangle\stackrel{\text{def}}{=}\int\int_{\Omega}P_{\infty} \mathrm{d}x\mathrm{d} z = 1,
\label{eqn:normalization_dimensionless}
\end{equation}
where $\Omega=\{(x, z): 0\leq x\leq 1, -\infty<z<\infty \}$ for a soft-channel and $\Omega=\{(x, z):0\leq x\leq L, z_-\leq z\leq z_+\}$ for a solid-channel, and the zero-flux condition,
\begin{subequations}
\begin{align}
J_{\infty}^z(x,\pm\infty)=0, \;\text{for a soft-channel},
\label{eqn:flux_dimensionless}\\
\epsilon^2\frac{\mathrm{d}z_{\pm}}{\mathrm{d}x}\left(\mathrm{Pe}-\frac{\partial P_{\infty}}{\partial x}\right)+\frac{\partial P_{\infty}}{\partial z}=0,\;\mathrm{at}\;z=z_{\pm}. \;\text{for a solid-channel}.
\end{align}
\end{subequations}

Once we obtain the asymptotic solution for the reduced probability distribution $P_{\infty}$,  we can calculate the average velocity along the channel by applying
macrotransport theory \cite{BrennerE1993},
\begin{equation}
U^*=\int\int_{\Omega}J_\infty^{x}\mathrm{d}x\mathrm{d}z,
\label{eqn:v}
\end{equation}
that is, the total flux in the $x$-direction averaged over a unit cell of the channel.
The effective dispersion coefficient $D^{*}$, can also be calculated from the asymptotic probability distribution $P_{\infty}$, via the so-called $B$-field,
which is defined by the following differential equation \cite{BrennerE1993},
\begin{equation}
\label{eqn:B}
\frac{\partial}{\partial z}\left(P_{\infty}\frac{\partial B}{\partial z}\right)-J^z_{\infty}\frac{\partial B}{\partial z}+\epsilon^{2}\left[\frac{\partial}{\partial x}\left(P_{\infty}\frac{\partial B}{\partial x}\right)-J^x_{\infty}\frac{\partial B}{\partial x}\right]=\epsilon^{2}P_{\infty}U^*.
\end{equation}
The boundary conditions for the $B$-field are
\begin{equation}
B(x=1,z)-B(x=0,z)=-1,\;\mathrm{and}
\end{equation}
\begin{subequations}
\begin{align}
P_\infty\frac{\partial B}{\partial z} \xrightarrow[{z \to \pm \infty}]{}0,\;\text{for a soft-channel}, \\
\epsilon^2\frac{\mathrm{d}z_{\pm}}{\mathrm{d}x}\frac{\partial B}{\partial x}=\frac{\partial B}{\partial z}\;\text{at}\;z=z_{\pm}(x),\;\text{for a solid-channel}.
\end{align}
\end{subequations}

Then, the effective diffusion coefficient is given by
\begin{equation}
\label{eqn:D}
D^{*}=\int\int_{\Omega}P_{\infty}\left[\left(
\frac{\partial B}{\partial x}\right)^{2}+\frac{1}{\epsilon^{2}}\left(\frac{\partial B}{\partial z}\right)^{2}\right]\mathrm{d}x\mathrm{d}z.
\end{equation}

\section{A narrow, slowly-varying soft-channel confined by a Parabolic Potential}
In this section, we consider a soft-channel in which particles are confined by a parabolic potential,
\begin{equation}
\bar{V}(\bar x,\bar z)=k_BT\pi\left(\frac{\bar{z}-\delta_z g(\bar{x}/L)}{a}\right)^2,
\label{eqn:potential_dimension}
\end{equation}
where $g(\bar x)$ is a periodic function. We have shown in previous work that the configuration integral,
\begin{equation}
\bar{I}(\bar{x}) = \int_{-\infty}^{\infty}e^{-\beta\bar{V}(\bar x,\bar z)}\mathrm{d}\bar z,
\end{equation}
with $\beta=1/k_BT$ plays a role analogous to the width of a solid-channel \cite{WangD2009}. Therefore, we shall call it the effective width of the soft-channel, which in this case is a constant for the potential in Eq. (\ref{eqn:potential_dimension}), $\bar{I}\left(\bar{x}\right)=a$.

\begin{figure}[!ht]
\centering
\includegraphics[width=5in]{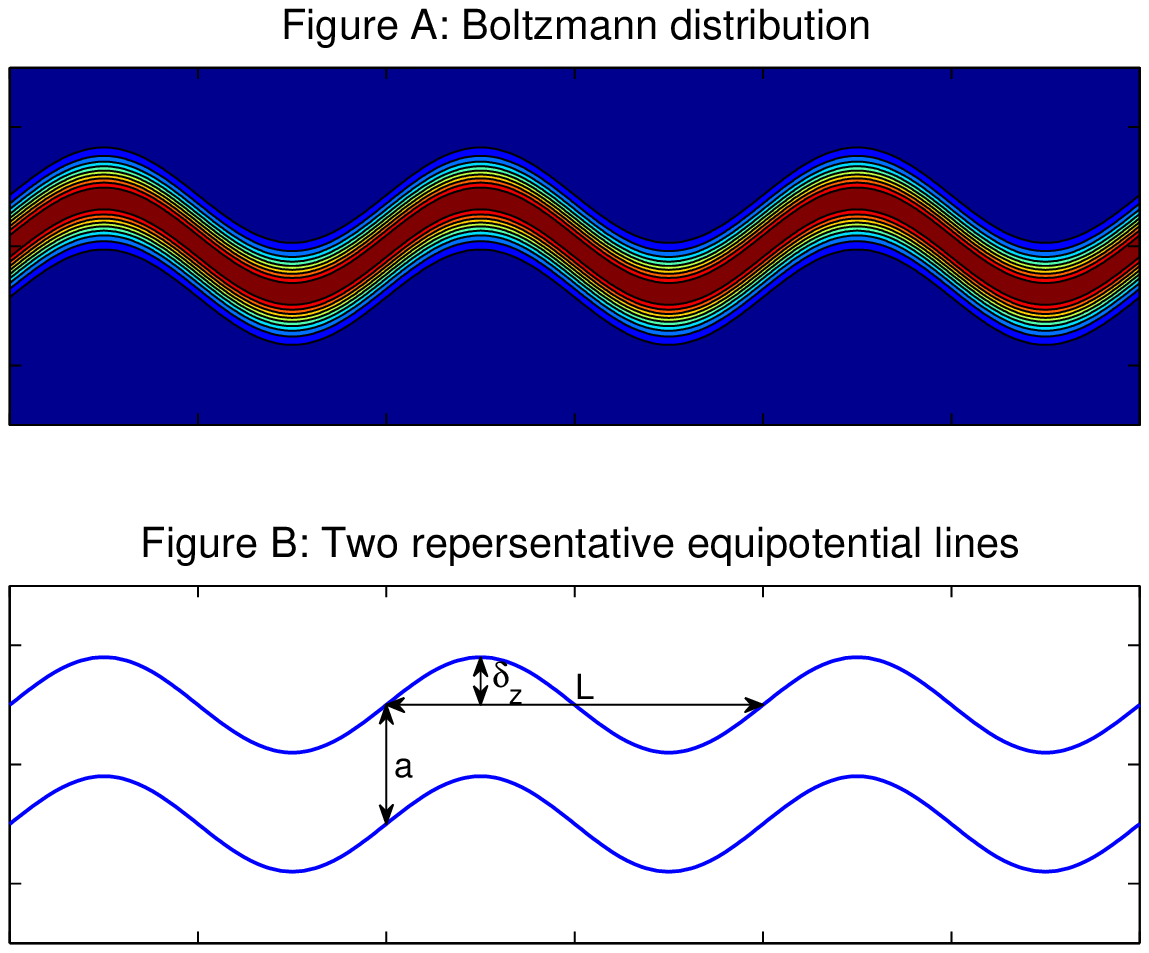}
\caption{(A) The Boltzmann distribution $\exp(-V)$ where $V(x,y)=\pi(z-\lambda g(x))^2$. (B) Schematic diagram of the curved channel confined by two equipotential lines $z=\lambda g(x)\pm 1/2$. The aspect ratio is $\epsilon=a/L$ and the ratio of the boundary amplitude to the width is $\lambda=\delta_z/a$.}
\label{fig:potentialcurve}
\end{figure}

The potential in dimensionless form is given by
\begin{equation}
V(x,z)=\pi\left(z-\lambda g(x)\right)^2,
\label{eqn:potential}
\end{equation}
where $\lambda=\delta_z/a$ is the ratio between the amplitude of the variations in the position of the centerline and the effective width of the soft-channel. The non-dimensional effective width of this soft-channel is $I(x)=1$. In equilibrium, the distribution of particles is given by the Boltzmann distribution, $\exp(-V)$, showed in Fig. \ref{fig:potentialcurve}(A). Fig. \ref{fig:potentialcurve}(B) shows a schematic diagram of a soft-channel whose boundaries are two equipotential lines. Note that particles are not strictly confined by these two boundaries. However, there is large probability that a particle is in the region inside two equipotential lines between which the distance is large. For example, if two soft-channel boundaries are equal potential lines $z=\lambda g(x)\pm1/2$, the probability that a particle is moving inside this soft-channel is about $79\%$ in equilibrium.

As discussed before, the aspect ratio is very small $\epsilon\ll1$, and the amplitude of the variation in the position of the centerline is of the same order as the width of the channel $\lambda\sim O(1)$. Therefore, we propose a solution for the stationary probability distribution in the form of a regular perturbation expansion in the small aspect ratio $\epsilon$,
\begin{equation}
P_{\infty}(x,z) \sim p_{0}+\epsilon^{2}p_{1}+\epsilon^{4}p_{2}+\cdots.
\end{equation}
The corresponding expansion for the probability flux is
\begin{equation}
{\bf J}_{\infty}(x,z) \sim {\bf J}_{0}+\epsilon^{2}{\bf J}_{1}+\epsilon^{4}{\bf J}_{2}+\cdots.
\end{equation}

At each order of the approximation, we first solve for the probability density $p_i(x,z)$, and we then calculate two important macroscopic transport properties: the average velocity given by Eq. (\ref{eqn:v}) and the effective diffusion coefficient given by Eq. (\ref{eqn:D}).

\subsection{Average velocity in a narrow, slowly-varying soft-channel}
The corresponding expansion of the average velocity in Eq. (\ref{eqn:v}) is
\begin{equation}
U^*_{soft}\sim u_0 + \epsilon^2 u_1 + \epsilon^4 u_2 + \cdots,
\end{equation}
where
\begin{subequations}
\begin{align}
u_0 &= \textrm{Pe}-\int_0^1\mathrm{d}x\int_{-\infty}^{\infty}\frac{\partial V}{\partial x}p_0\mathrm{d}z,\label{eqn:u0}\\
u_i &= -\int_0^1\mathrm{d}x\int_{-\infty}^{\infty}\frac{\partial V}{\partial x}p_i\mathrm{d}z, \quad\textrm{for}\quad i=1,2,3\label{eqn:ui}\cdots.
\end{align}
\end{subequations}
On the other hand, integrating both sides of Eq. (\ref{eqn:governing}) over the cross section and, applying the far-field conditions, we obtain
\begin{equation}
\frac{\mathrm{d}}{\mathrm{d}x}\left\{\int_{-\infty}^{\infty}\left[\left(\text{Pe}-\frac{\partial V}{\partial x}\right)P_{\infty}-\frac{\partial P_{\infty}}{\partial x}\right]\mathrm{d}z\right\}=0.
\end{equation}
This shows that the total flux in the $x$-direction $\bar{J}^x$ (the quantity inside the curly brackets) is, in steady state, constant along the channel. Furthermore, given the definition of the average velocity, $U^*_{soft}=\int_0^1\bar{J}^x\mathrm{d}x$, we have that $\bar{J}^x=U^*_{soft}$. Therefore,
\begin{equation}
u_i=\int_{-\infty}^{\infty}\left[\left(\text{Pe}-\frac{\partial V}{\partial x}\right)p_i-\frac{\partial p_i}{\partial x}\right]\mathrm{d}z.
\label{eqn:ui_new}
\end{equation}
This is a solvability condition for $p_i(x,z)$, which can also be derived from next order governing equation.
In what follows, Eq. (\ref{eqn:u0}) and Eq. (\ref{eqn:ui}) are used to calculate the average velocity $u_i$.
We shall show that it is possible to obtain $u_i$ by first finding  $p_i(x,z)$ up to an unknown function of $x$.
Then, the unknown part  of $p_i(x,z)$ is determined by means of Eq. (\ref{eqn:ui_new}).

In order to calculate the average velocity $u_i$, we first need to calculate the probability density $p_i$. Substituting the expansion of $P_{\infty}$ into Eq. (\ref{eqn:governing}), we determine the leading order governing equation,
\begin{equation}
\frac{\partial}{\partial z}\left(-\frac{\partial V}{\partial z}p_{0}-\frac{\partial p_{0}}{\partial z}\right)=\frac{\partial J_{0}^{z}}{\partial z}=0.
\label{eqn:governing_0}
\end{equation}

The corresponding leading order boundary and normalization conditions, derived from Eqs. (\ref{eqn:normalization}-\ref{eqn:flux}), are
\begin{subequations}
\begin{align}
J_{0}^{z}\left(x,\pm\infty\right) &= 0, \\
p_{0}\left(0,z\right) &= p_{0}\left(1,z\right), \\
\left<p_{0}\right> &= 1.
\end{align}
\end{subequations}
Eq. (\ref{eqn:governing_0}) shows that the flux $J_{0}^{z}$ is independent of $z$, which in combination with the zero flux condition for $z\rightarrow \pm\infty$, implies that $J_{0}^{z}=0$. Then, the leading order solution of the probability density is
\begin{equation}
p_{0}(x,z)=f_0(x)e^{-V(x,z)},
\label{eqn:p0}
\end{equation}
where $f_0(x)$ is an unknown function which satisfies the periodicity and normalization conditions. As we mentioned before, without knowing the explicit solution of $f_0(x)$, we can still calculate the leading order average velocity according to Eq. (\ref{eqn:u0}) by taking advantage of the relation $\frac{\partial V}{\partial x}=-\lambda\frac{\mathrm{d}g}{\mathrm{d}x}\frac{\partial V}{\partial z}$,
\begin{equation}
u_0=\text{Pe}.
\end{equation}

The explicit solution of $f_0(x)$ can be determined from Eq. (\ref{eqn:ui_new}),
\begin{equation}
\text{Pe}=\int_{-\infty}^{\infty}\left[\left(\text{Pe}-\frac{\partial V}{\partial x}\right)p_0-\frac{\partial p_0}{\partial x}\right]\mathrm{d}z.
\end{equation}
By substituting the solution of $p_0$ and evaluating the integral, we obtain,
\begin{equation}
\text{Pe}=\left(\text{Pe}f_0-\frac{\mathrm{d}f_0}{\mathrm{d}x}\right)I(x),
\end{equation}
where $I(x)=\int_{-\infty}^{\infty}e^{-V}\mathrm{d}z=1$. This leads to $f_0(x)=1$. Thus the leading order term of the probability density is
\begin{equation}
p_0(x,z)=e^{-V},
\end{equation}
i.e., the Boltzmann distribution. This means that a small variation in the centerline of the channel does not affect the leading order solution. Therefore, we seek higher order corrections $p_n\; \mathrm{for}\;n=1,2,3,\cdots$, which satisfy
\begin{equation}
\frac{\partial}{\partial z}\left(\frac{\partial V}{\partial z}p_{n}+\frac{\partial p_{n}}{\partial z}\right)=\frac{\partial}{\partial x}\left[\left(\textrm{Pe}-\frac{\partial V}{\partial x}\right)p_{n-1}-\frac{\partial p_{n-1}}{\partial x}\right].
\label{eqn:governing_1}
\end{equation}
The corresponding boundary and normalization conditions are
\begin{subequations}
\begin{align}
J_{n}^{z}\left(x,\pm\infty\right) &= 0, \\
p_{n}\left(0,z\right) &= p_{n}\left(1,z\right),\\
\left<p_{n}\right> &= 0. \label{eqn:BCs_1}
\end{align}
\end{subequations}

Substituting $p_{0}=e^{-V}$ into Eq. (\ref{eqn:governing_1}) for $n=1$, we obtain
%

\begin{equation}
\frac{\partial}{\partial z}\left(\frac{\partial V}{\partial z}p_1+\frac{\partial p_1}{\partial z}\right)=\textrm{Pe}\frac{\partial p_0}{\partial x}.
\end{equation}
Integrating both sides of the equation twice with respect to $z$, we find the solution of the probability density term at $O(\epsilon^2)$,
\begin{equation}
p_1(x,z)=f_1(x,z)e^{-V(x,z)},
\end{equation}
where $f_1(x,z)=-\lambda\textrm{Pe}\frac{\mathrm{d}g}{\mathrm{d}x}\,z+C_1(x)$. The function $C_1(x)$ can be determined from Eq. (\ref{eqn:ui_new}).


Analogous to the leading order term, we can evaluate $u_1$ without knowing $C_1(x)$,
\begin{equation}
\begin{split}
u_1&=-\int_0^1\mathrm{d}x\int_{-\infty}^{\infty}\frac{\partial V}{\partial x}p_1\mathrm{d}z\\
&=-\int_0^1\mathrm{d}x\int_{-\infty}^{\infty}\left(-\lambda\frac{\mathrm{d}g}{\mathrm{d}x}\frac{\partial V}{\partial z}\right)f_1(x,z)e^{-V}\mathrm{d}z\\
&=\int_0^1\mathrm{d}x\int_{-\infty}^{\infty}\left(\lambda\frac{\mathrm{d}g}{\mathrm{d}x}\right)
\frac{\partial f_1}{\partial z}e^{-V}\mathrm{d}z\\
&=-\lambda^2\textrm{Pe}\int_0^1\left(\frac{\mathrm{d}g}{\mathrm{d}x}\right)^2\mathrm{d}x.
\end{split}
\end{equation}

Continuing with the same approach, we can solve the problem at $O(\epsilon^4)$. The probability density is
\begin{equation}
p_2(x,z)=f_{2}(x,z)e^{-V},
\end{equation}
where
\begin{equation}
f_{2}(x,z)=\frac{\textrm{Pe}\lambda^2}{2}\frac{\mathrm{d}g}{\mathrm{d}x}\left(\textrm{Pe}
\frac{\mathrm{d}g}{\mathrm{d}x}- \frac{\mathrm{d}^2g}{\mathrm{d}x^2}\right)\left(z-\lambda g(x)\right)^2 +\frac{\textrm{Pe}\lambda}{2\pi}\left(\textrm{Pe}\frac{\mathrm{d}^2g}{\mathrm{d}x^2}-
\frac{\mathrm{d}^3g}{\mathrm{d}x^3}\right)z-u_1\lambda\frac{\mathrm{d}g}{\mathrm{d}x}z+ C_2(x).
\end{equation}
The corresponding average velocity is
\begin{equation}
u_2=\frac{\textrm{Pe}\lambda^2}{2\pi}\int_0^1\left(\frac{\mathrm{d}^2g}{\mathrm{d}x^2}\right)^2dx+\textrm{Pe}
\lambda^4\left(\int_0^1\left(\frac{\mathrm{d}g}{\mathrm{d}x}\right)^2dx\right)^2.
\end{equation}

In summary, the average velocity up to $O(\epsilon^4)$ is
\begin{equation}
\begin{split}
U^*_{soft}\sim\textrm{Pe}-\epsilon^2\lambda^2\textrm{Pe}\int_0^1\left(\frac{\mathrm{d}g}{\mathrm{d}x}\right)^2\mathrm{d}x+ \\
\epsilon^4\textrm{Pe}\left[\frac{\lambda^2}{2\pi}\int_0^1\left(\frac{\mathrm{d}^2g}{\mathrm{d}x^2}\right)^2\mathrm{d}x+\lambda^4\left(
\int_0^1\left(\frac{\mathrm{d}g}{\mathrm{d}x}\right)^2\mathrm{d}x\right)^2\right]+O(\epsilon^6).
\label{eqn:U*_soft}
\end{split}
\end{equation}
Note that, to this order of the approximation, the average velocity depends linearly on the P\'eclet number. Equivalently, the normalized mobility is constant and independent of the P\'eclet number,
\begin{equation}
\begin{split}
\mu_{soft}=\frac{U^*}{\text{Pe}}\sim1-\epsilon^2\lambda^2\int_0^1\left(\frac{\mathrm{d}g}{\mathrm{d}x}\right)^2\mathrm{d}x
+\\
\epsilon^4\left[\frac{\lambda^2}{2\pi}\int_0^1\left(\frac{\mathrm{d}^2g}{\mathrm{d}x^2}\right)^2\mathrm{d}x+\lambda^4\left(
\int_0^1\left(\frac{\mathrm{d}g}{\mathrm{d}x}\right)^2\mathrm{d}x\right)^2\right]+O(\epsilon^6).
\label{eqn:mu_soft}
\end{split}
\end{equation}

\subsection{Effective diffusion coefficient in a narrow, slowly-varying soft-channel }
\label{section:D_epsilon}
In order to calculate the effective diffusion coefficient $D^*_{soft}$, we need to solve the $B$-field in Eq. (\ref{eqn:B}). Asymptotic expansions are proposed in the following form:
\begin{equation}
B(x,z) \sim B_0 + \epsilon^2 B_1 + \epsilon^4 B_2+\cdots,
\end{equation}
\begin{equation}
D^*_{soft} \sim D_{0}+\epsilon^{2}D_{1}+\epsilon^{4}D_{2}+\cdots,
\end{equation}
where
\begin{eqnarray}
D_0 &=&\int_0^1\mathrm{d}x\int_{-\infty}^{\infty}p_0\left(\frac{\partial B_0}{\partial x}\right)^2\mathrm{d}z,\\
D_1 &=&\int_0^1\mathrm{d}x\int_{-\infty}^{\infty}\left\{p_0\left[2\frac{\partial B_0}{\partial x}\frac{\partial B_1}{\partial x}+\left(\frac{\partial B_1}{\partial z}\right)^2\right]+p_1\left(\frac{\partial B_0}{\partial x}\right)^2\right\}\mathrm{d}z.
\end{eqnarray}

The leading order governing equation derived from Eq. (\ref{eqn:B}) is, after simplifications,
\begin{equation}
\frac{\partial}{\partial z}\left(p_0\frac{\partial B_0}{\partial z}\right)=0,
\label{eqn:B0_full}
\end{equation}
and the boundary conditions at $O(1)$ are
\begin{eqnarray}
p_0\frac{\partial B_0}{\partial z} \xrightarrow[{z \to \pm \infty}]{}0, \\
B_0(1,z)-B_0(0,z)=-1.
\end{eqnarray}
Therefore $B_0$ is a function of $x$ only. The exact solution, up to an arbitrary additive constant, can be derived by integrating the governing equation over the cross section at the next order in the expansion \cite{WangD2009}, that is
\begin{equation}
\frac{\mathrm{d}}{\mathrm{d}x}\int_{-\infty}^{\infty}\left(P_{\infty}\frac{\partial B}{\partial x}-BJ^x_{\infty}\right)\mathrm{d}z=U^*_{soft}\int_{-\infty}^{\infty}P_{\infty}\mathrm{d}z.
\label{eqn:B_integral}
\end{equation}
The leading order of this equation can be simplified to obtain
\begin{equation}
\frac{\mathrm{d}^2B_0}{\mathrm{d}x^2}-\textrm{Pe}\frac{\mathrm{d}B_0}{\mathrm{d}x}=\textrm{Pe},
\end{equation}
which gives
\begin{equation}
\frac{dB_0}{dx}=-1.
\end{equation}
Then, the leading order of the effective diffusion coefficient is
\begin{equation}
D_0=\int_0^1\mathrm{d}x\int_{-\infty}^{\infty}p_0\left(\frac{\mathrm{d}B_0}{\mathrm{d}x}\right)^2\mathrm{d}z=1.
\end{equation}
This result is consistent with the leading order term for the average velocity, which was also not affected by the variation in the position of the channel centerline.

The governing equation at $O(\epsilon^2)$ is also derived from Eq. (\ref{eqn:B}). After some simplifications, we obtain
\begin{equation}
\frac{\partial}{\partial z}\left(p_0\frac{\partial B_1}{\partial z}\right)+\frac{\partial}{\partial x}\left(p_0\frac{\partial B_0}{\partial x}\right)-\left[\left(\textrm{Pe}-\frac{\partial V}{\partial x}\right)p_0-\frac{\partial p_0}{\partial x}\right]\frac{\partial B_0}{\partial x}=p_0u_0,
\label{eqn:B0}
\end{equation}
with the boundary conditions
\begin{eqnarray}
p_0\frac{\partial B_1}{\partial z} \xrightarrow[{z \to \pm \infty}]{}0, \\
B_1(1,z)-B_1(0,z)=0.
\end{eqnarray}

Substituting the functions of $B_0$, $p_0$, and $u_0$ into Eq. (\ref{eqn:B0}) and integrating twice with respect to $z$, we obtain
\begin{equation}
B_1(x,z)=-\lambda\frac{\mathrm{d}g}{\mathrm{d}x} z+K_{1}(x),
\end{equation}
with the condition $K_{1}(0)=K_{1}(1)$ derived from the condition $B_1(0,z)=B_1(1,z)$. After some simplifications, the effective diffusion coefficient at $O(\epsilon^2)$ is given by
\begin{eqnarray}
D_1&=&\int_0^1dx \int_{-\infty}^{\infty}\left\{ p_0\left[2\frac{\partial B_0}{\partial x}\frac{\partial B_1}{\partial x}+\left(\frac{\partial B_1}{\partial z}\right)^2\right]+p_1\left(\frac{\partial B_0}{\partial x}\right)^2 \right\}\mathrm{d}z\nonumber\\
&=&-\lambda^2\int_0^1\left(\frac{\mathrm{d}g}{\mathrm{d}x}\right)^2dx.
\end{eqnarray}

Therefore, the effective diffusion coefficient up to $O(\epsilon^2)$ is given by
\begin{equation}
D^{*}_{soft}\sim 1-\epsilon^2\lambda^2\int_{0}^{1}\left(\frac{\mathrm{d}g}{\mathrm{d}x}\right)^2\mathrm{d}z.
\end{equation}
This recovers the Einstein-Smoluchowski relation in the dimensionless form, which is $D^{*}_{soft}=\mu^{*}_{soft}$.

\section{Transport in a narrow, slowly-varying solid-channel}
In this section, we consider a solid-channel with upper and lower walls described by $z_{\pm}=\lambda g(x)\pm 1/2$. The function $\lambda g(x)$ corresponds to the centerline of the channel, and it is periodic $g(0)=g(1)$. The channel width, $w(x)=z_+-z_-=1$, is equal to the effective width $I(x)=1$ of the soft-channel considered in the previous section. The dimensionless governing equation, with $V(x,z)=0$, becomes
\begin{equation}
\epsilon^2\frac{\partial}{\partial x}\left(\textrm{Pe}P_\infty-\frac{\partial P_{\infty}}{\partial x}\right)-\frac{\partial^2 P_{\infty}}{\partial z^2}=0.
\label{eqn:Psolid}
\end{equation}
The periodic boundary condition in $x$, the zero-flux condition at the boundaries, and the normalization condition are given by
\begin{subequations}
\begin{align}
P_{\infty}(0,z)&=P_{\infty}(1,z),\\
-\epsilon^2\lambda g'(x)\left(\textrm{Pe}P_\infty-\frac{\partial P_{\infty}}{\partial x}\right)&=\frac{\partial P_{\infty}}{\partial z},\; \text{at}\; z=\lambda g(x)\pm \frac{1}{2},\\
\int_0^1\mathrm{d}x\int_{\lambda g(x)-1/2}^{\lambda g(x)+1/2} P_{\infty}\mathrm{d}z&=1.
\end{align}
\end{subequations}

\subsection{Average velocity in a narrow, slowly-varying solid-channel}
Analogous to the analysis presented for soft-channels, we focus on the limiting case of $\epsilon\ll 1$ and $\lambda\sim O(1)$. First, we propose a solution in the form of an asymptotic expansion of $P_\infty$,
\begin{equation}
P_\infty\sim \rho_0(x,z)+\epsilon^2\rho_1(x,z)+\epsilon^4\rho_2(x,z)+\cdots.
\end{equation}
after the probability density $P_\infty$ is determined, the average velocity can be evaluated by
\begin{equation}
\begin{split}
U^*_{solid} &= \int_0^1\mathrm{d}x\int_{\lambda g(x)-1/2}^{\lambda g(x)+1/2}\left(\text{Pe}P_{\infty}-\frac{\partial P_{\infty}}{\partial x}\right)\mathrm{d}z\\
&\sim v_0+\epsilon^2 v_1 + \epsilon^4 v_2 + \cdots,
\end{split}
\end{equation}
where
\begin{subequations}
\begin{align}
v_0&=\text{Pe}-\int_0^1\mathrm{d}x\int_{\lambda g(x)-1/2}^{\lambda g(x)+1/2}\frac{\partial \rho_0}{\partial x}\mathrm{d}z,\\
v_i&=-\int_0^1\mathrm{d}x\int_{\lambda g(x)-1/2}^{\lambda g(x)+1/2}\frac{\partial \rho_i}{\partial x}\mathrm{d}z, \; \text{for} \;i=1,2,3,\cdots.
\end{align}
\end{subequations}

On the other hand, integrating both sides of Eq. (\ref{eqn:Psolid}) and applying the zero-flux boundary conditions, we obtain
\begin{equation}
\frac{\mathrm{d}}{\mathrm{d}x}\left[\int_{\lambda g(x)-1/2}^{\lambda g(x)+1/2}\left(\textrm{Pe}P_\infty-\frac{\partial P_{\infty}}{\partial x}\right)\mathrm{d}z\right]=0.
\end{equation}
Therefore, as expected, the quantity inside square brackets, which is the total flux in $x$-direction, is constant along the channel. Since the integral of the total flux in $x$-direction is the average velocity, the we obtain,
\begin{equation}
U^*_{solid}=
\int_{\lambda g(x)-1/2}^{\lambda g(x)+1/2}\left(\textrm{Pe}P_\infty-\frac{\partial P_{\infty}}{\partial x}\right)\mathrm{d}z,
\end{equation}
or
\begin{equation}
\begin{split}
v_i&=\int_{\lambda g(x)-1/2}^{\lambda g(x)+1/2}\left(\textrm{Pe}\rho_i-\frac{\partial \rho_i}{\partial x}\right)\mathrm{d}z\\
&=\text{Pe}\bar{\rho}_i-\left(\frac{\mathrm{d}\bar{\rho}_i}{\mathrm{d}x}-\lambda\frac{\mathrm{d}g}{\mathrm{d}x}
\rho_i|_{z=\lambda g(x)+1/2}+\lambda\frac{\mathrm{d}g}{\mathrm{d}x}
\rho_i|_{z=\lambda g(x)-1/2}\right),
\label{eqn:vi0}
\end{split}
\end{equation}
where $\bar{\rho}_i$ is the marginal probability density, $\bar{\rho}_i=\int_{\lambda g(x)-1/2}^{\lambda g(x)+1/2}\rho_i\mathrm{d}z$. Then, integrating both sides of the equation above with respect to $x$ from $0$ to $1$, the first term on the right hand side of the equation cancels for $i\geq1$, due to the normalization condition; the second term of the right hand side is also identically zero, due to the periodicity in $x$. Therefore, we obtain an alternative expression for the average velocity,
\begin{equation}
v_i = \int_0^1\lambda\frac{\mathrm{d}g}{\mathrm{d}x}
\left(\rho_i|_{z=\lambda g(x)+1/2}-\rho_i|_{z=\lambda g(x)-1/2}\right)\mathrm{d}x,\;\text{for}\; i=1,2,3,\cdots.
\label{eqn:vi}
\end{equation}
This equation shows that the average velocity is completely determined by the probability density on the upper and lower boundaries. We shall use this expression to calculate the average velocity.

It is straightforward to show that the leading order term for the probability density is uniform, $\rho_0=1$ inside the channel. The corresponding leading order contribution to the average velocity is $v_0=\text{Pe}$. The higher order terms of the probability density are governed by
\begin{equation}
\frac{\partial ^2 \rho_i}{\partial z^2}=\frac{\partial}{\partial x}\left(\text{Pe}\rho_{i-1}-\frac{\partial \rho_{i-1}}{\partial x}\right),\;\text{for}\; i=1, 2, 3, \cdots,
\label{eqn:rho}
\end{equation}
and satisfy both the zero-flux boundary condition,
\begin{equation}
\frac{\partial \rho_i}{\partial z}=-\lambda \frac{\mathrm{d}g}{\mathrm{d}x}\left(\text{Pe}\rho_{i-1}-\frac{\partial \rho_{i-1}}{\partial x}\right),\;\text{for}\; i=1, 2, 3, \cdots,
\label{eqn:noflux_rho}
\end{equation}
as well as the normalization condition,
\begin{equation}
\int_0^1\mathrm{d}x\int_{\lambda g(x)-1/2}^{\lambda g(x)+1/2}\rho_i\mathrm{d}z=0,\;\text{for}\; i=1, 2, 3, \cdots.
\label{eqn:normalization_rho}
\end{equation}
Substituting $\rho_0$ into Eq. (\ref{eqn:rho}), we obtain $\partial^2\rho_1/\partial z^2=0$, which corresponds to a solution of the form
\begin{equation}
\rho_1(x,z)=a^1_1(x)z+a^1_0(x),
\end{equation}
where $a^1_1(x)$ is determined by the zero-flux condition,
\begin{equation}
a^1_1(x)=-\lambda\text{Pe}\frac{\mathrm{d}g}{\mathrm{d}x},
\end{equation}
 and the normalization condition implies $\int_0^1a^1_0(x)\mathrm{d}x=0$. The governing equation for $a^1_0(x)$ can be derived by integrating the $O(\epsilon^4)$ terms of the governing equation with respect to $z$ over the cross section,
\begin{equation}
\mathrm{Pe}a_0^1-\frac{\mathrm{d}a_0^1}{\mathrm{d}x}=v_1-\left(\mathrm{Pe}a_1^1
-\frac{\mathrm{d}a_1^1}{\mathrm{d}x}\right)\lambda g(x).
\end{equation}
However, it is not necessary to determine $a^1_0(x)$ for calculating the average velocity. In fact, from Eq. (\ref{eqn:vi}) we obtain
\begin{equation}
v_1=-\lambda^2\text{Pe}\int_0^1\left(\frac{\mathrm{d}g}{\mathrm{d}x}\right)^2\mathrm{d}x.
\end{equation}

Before calculating higher order contributions, we present the general procedure to calculate $\rho_i$ for $i=1,2,3,\cdots$. Since $\rho_1$ is a first order polynomial in terms of $z$, by induction, we propose $\rho_i$ to be a polynomial of degree $(2i-1)$ in terms of $z$,
\begin{equation}
\rho_i(x,z)=a^i_{2i-1}(x)\frac{z^{2i-1}}{(2i-1)!}+a^i_{2i-2}(x)\frac{z^{2i-2}}{(2i-2)!}+
\cdots+a^i_1(x)\frac{z}{1!}+a^i_0(x).
\end{equation}
The coefficient $a^i_0(x)$ can be determined from the normalization condition for the next order term in the asymptotic solution. The coefficient $a^i_1(x)$ can be determined by the zero-flux condition in Eq. (\ref{eqn:noflux_rho}), and all other coefficients $a^i_j(x)$ can be determined from the coefficients of the lower order term in the asymptotic solution,
\begin{equation*}
a^i_j(x)=\frac{\mathrm{d}}{\mathrm{d} x}\left[\text{Pe}a^{i-1}_{j-2}(x)-\frac{\mathrm{d} a^{i-1}_{j-2}(x)}{\mathrm{d} x}\right], \;\text{for}\; j=2, 3,\cdots,2i-1.
\end{equation*}

In principle, higher order terms can be obtained using the proposed procedure repeatedly. Here, for simplicity, we only show the results up to $O(\epsilon^4)$,
\begin{equation}
\rho_2(x,z)=a^2_3(x)\frac{z^3}{3!}+a^2_2(x)\frac{z^2}{2!}+a^2_1(x)\frac{z}{1!}+a^2_0(x),
\end{equation}
where $a^2_3(x)$ and $a^2_2(x)$ are derived from $a^1_1(x)$ and $a^1_0(x)$, respectively,
\begin{equation}
\begin{split}
a^2_3(x)&=\frac{\mathrm{d}}{\mathrm{d} x}\left[\text{Pe}a^1_1(x)-\frac{\mathrm{d} a^1_1(x)}{\mathrm{d} x}\right]\\
&=-\lambda\mathrm{Pe}\left[\mathrm{Pe}\frac{\mathrm{d}^2g}{\mathrm{d}x^2}-
\frac{\mathrm{d}^3g}{\mathrm{d}x^3}\right],
\end{split}
\end{equation}
\begin{equation}
\begin{split}
a^2_2(x)&=\frac{\mathrm{d}}{\mathrm{d} x}\left[\text{Pe}a^1_0(x)-\frac{\mathrm{d} a^1_0(x)}{\mathrm{d} x}\right]\\
&=\lambda^2\mathrm{Pe}\frac{\mathrm{d}}{\mathrm{d}x}\left[g(x)
\left(\mathrm{Pe}\frac{\mathrm{d}g}{\mathrm{d}x}-\frac{\mathrm{d}^2g}{\mathrm{d}x^2}\right)\right].
\end{split}
\end{equation}
Then, $a_1^2(x)$ is determined by the zero-flux boundary condition,
\begin{equation}
a^2_1(x)=-\lambda\frac{\mathrm{d}g}{\mathrm{d}x}\left(\text{Pe}\rho_1-\frac{\mathrm{\partial \rho_1}}{\partial x}\right)\bigg|_{z=\lambda g(x)+1/2}
-\left(a^2_3(x)\frac{z^2}{2!}+a^2_2(x)\frac{z}{1!}\right)\bigg|_{z=\lambda g(x)+1/2}.
\end{equation}
Finally, the next order correction to the average velocity, $v_2$, is evaluated from Eq. (\ref{eqn:vi}),
\begin{equation}
v_2=\frac{1}{12}\text{Pe}\lambda^2\int_0^1\left(\frac{\mathrm{d}^2g}{\mathrm{d}x^2}\right)^2\mathrm{d}x
+\text{Pe}\lambda^4\left[\int_0^1\left(\frac{\mathrm{d}g}{\mathrm{d}x}\right)^2\mathrm{d}x\right]^2.
\end{equation}

\begin{figure}[!ht]
\centering
\includegraphics[width=5in]{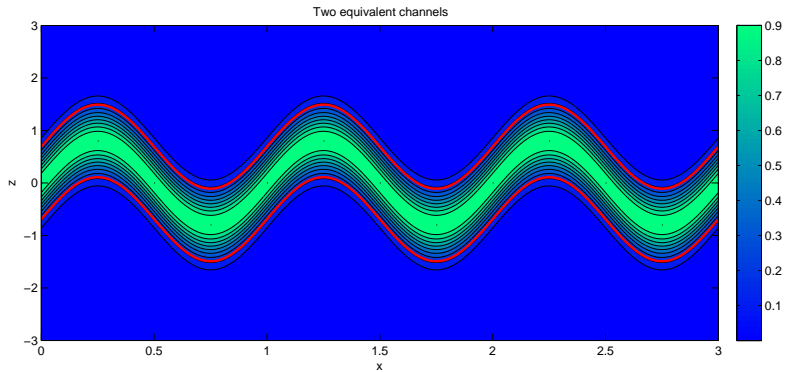}
\caption{Two equivalent channels: the solid-channel is bounded by two red lines $z_{\pm}=\lambda g(x)\pm\frac{\sqrt{6/\pi}}{2}$ and the concentration in the soft-channel is given by the Boltzmann distribution $e^{-\pi\left(z-\lambda g(x)\right)^2}$ where $\epsilon=0.1$ and $\lambda=0.8$.}
\label{fig:TwoEquivalentChannel}
\end{figure}

Adding these contributions, the dimensionless mobility up to $O(\epsilon^4)$ is given by
\begin{equation}
\mu_{solid}=\frac{U^*_{solid}}{\text{Pe}}\sim1-\epsilon^2\lambda^2\int_0^1\left(\frac{\mathrm{d}g}
{\mathrm{d}x}\right)^2\mathrm{d}x+\epsilon^4\left\{\frac{1}{12}\lambda^2\int_0^1\left(\frac{\mathrm{d}^2g}
{\mathrm{d}x^2}\right)^2\mathrm{d}x+\lambda^4\left[\int_0^1\left(\frac{\mathrm{d}g}
{\mathrm{d}x}\right)^2\mathrm{d}x\right]^2\right\}.
\label{eqn:mu_solid}
\end{equation}
First, we note that the mobility up to $O(\epsilon^4)$ is independent of the P\'eclet number, as in the case of soft-channels. Comparing this mobility, $\mu_{solid}$, with $\mu_{soft}$ obtained for the soft-channel, it is clear that the resistance of the solid-channel is higher than that of the soft-channel. Specifically, the effect of the second derivative of the position of the centerline of the solid channel is nearly one half of that in a soft-channel. This conclusion is different from that obtained from the transport in a weakly-corrugated symmetric channel with a varying width, in which the resistance of the solid-channel could be smaller than that of the soft-channel for small P\'eclet numbers \cite{WangD2010}. However we can make $(\epsilon^4\frac{1}{12}\lambda^2)_{solid}=(\epsilon^4\frac{1}{2\pi}\lambda^2)_{soft}$ by letting $\epsilon_{soft}=\sqrt{\pi/6}\epsilon_{solid}$ and $\lambda_{soft}=\sqrt{6/\pi}\lambda_{solid}$. For example, in a dimensional case, we can increase the solid-channel width from $w(x)=b$ to $b\sqrt{6/\pi}$ or reducing the soft-channel width from $I(x)=b$ to $b/\sqrt{6/\pi}$. Thus, if $I(x)=w(x)/\sqrt{6/\pi}$, the average velocity in both soft and solid-channels are the same up to $O(\epsilon^4)$.
The Fig. \ref{fig:TwoEquivalentChannel} shows two equivalent channels where the solid-channel is bounded by two red lines $z_{\pm}=\lambda g(x)\pm\frac{\sqrt{6/\pi}}{2}$ and the soft-channel is represented by the concentration of particles, that is the Boltzmann distribution $e^{-\pi\left(z-\lambda g(x)\right)^2}$.

\subsection{Effective diffusion coefficient in a narrow, slowly-varying solid-channel}
In order to calculate the effective diffusion coefficient, $D^*_{solid}$, we need to solve the $B$-field equation,
\begin{equation}
\frac{\partial}{\partial z}\left(P_{\infty}\frac{\partial B}{\partial z}\right)
-\left(-\frac{\partial P_{\infty}}{\partial z}\right)\frac{\partial B}{\partial z}
+\epsilon^2\left[\frac{\partial}{\partial x}\left(P_{\infty}\frac{\partial B}{\partial x}\right)-\left(\text{Pe}P_{\infty}-\frac{\partial P_{\infty}}{\partial x}\right)\frac{\partial B}{\partial x}\right]=\epsilon^2P_{\infty}U^*_{solid},
\end{equation}
with boundary conditions,
\begin{subequations}
\begin{align}
\frac{\partial B}{\partial z}=\epsilon^2\lambda\frac{dg}{dx}\frac{\partial B}{\partial x},\;\text{at}\;z=\lambda g(x)\pm \frac{1}{2},\\
B(1,z)-B(0,z)=-1.
\end{align}
\end{subequations}
Proposing then an asymptotic expansion for the $B$-field,
\begin{equation}
B(x,z)\sim \mathcal{B}_0(x,z)+\epsilon^2\mathcal{B}_1(x,z)+\epsilon^4\mathcal{B}_2(x,z)+\cdots,
\end{equation}
and solving for $\mathcal{B}_i(x,z)$, the effective diffusion coefficient is given by
\begin{equation}
\begin{split}
D^*_{solid}&=\int_0^1\mathrm{d}x\int_{\lambda g(x)-1/2}^{\lambda g(x)+1/2}\left[\left(\frac{\partial B}{\partial x}\right)^2+\frac{1}{\epsilon^2}\left(\frac{\partial B}{\partial z}\right)^2\right]\mathrm{d}z\\
&\sim \mathcal{D}_0+\epsilon^2\mathcal{D}_1+\epsilon^4\mathcal{D}_2+\cdots,
\end{split}
\end{equation}
where
\begin{eqnarray}
\mathcal{D}_0 &=&\int_0^1\mathrm{d}x\int_{\lambda g(x)-1/2}^{\lambda g(x)+1/2}\rho_0\left(\frac{\partial \mathcal{B}_0}{\partial x}\right)^2\mathrm{d}z,\\
\mathcal{D}_1 &=&\int_0^1\mathrm{d}x\int_{\lambda g(x)-1/2}^{\lambda g(x)+1/2}\left\{\rho_0\left[2\frac{\partial \mathcal{B}_0}{\partial x}\frac{\partial \mathcal{B}_1}{\partial x}+\left(\frac{\partial \mathcal{B}_1}{\partial z}\right)^2\right]+\rho_1\left(\frac{\partial \mathcal{B}_0}{\partial x}\right)^2\right\}\mathrm{d}z.
\end{eqnarray}

It is straightforward to calculate the effective diffusion coefficient up to $O(\epsilon^2)$,
\begin{equation}
D^*_{solid}\sim 1-\epsilon^2\lambda^2\int_0^1\left(\frac{\mathrm{d}g}{\mathrm{d}x}\right)^2\mathrm{d}x.
\end{equation}
We have shown before that the corrections to the mobility are independent of the P\'eclet number. Therefore, as expected, the Einstein-Smoluchowski relation $D^*_{solid}=\mu^*_{solid}$ is also valid in the case of solid-channels.

\section{Discussion}
Physically, the dimensional average velocity discussed above can be expressed as the ratio between the distance traveled in the longitudinal direction and the nominal holdup time within the curved channel \cite{RushDBK2002},
\begin{equation}
\bar{U}^*=\frac{\text{longitudinal distance traveled}}{\text{nominal holdup time}}=\frac{L}{t^*}.
\end{equation}
The nominal holdup time is the average transit time between the two ends of the channel, separated a distance L, and it is given by the channel arclength divided by the velocity along the curved channel. The driving force $F$ is constant along the $\bar{x}$-direction. Then, to calculate the velocity along the channel centerline we first write its tangent, $(\mathrm{d}\bar{x}, \mathrm{d}\bar{z})$ in the dimensional form. Then, the arclength is $\mathrm{d}s=\sqrt{\mathrm{d}\bar{x}^2+\mathrm{d}\bar{z}^2}$ and the velocity along the centerline is $\frac{\mathrm{d}\bar{x}}{\mathrm{d}s}\frac{F}{\eta}$. Therefore, the nominal holdup time is
\begin{equation}
t^*=\int_0^L\frac{ds}{\frac{\mathrm{d}\bar{x}}{\mathrm{d}s}\frac{F}{\eta}}
=\frac{\eta}{F}\int_0^L\left[1+\left(\frac{\mathrm{d}\bar{z}}
{\mathrm{d}\bar{x}}\right)^2\right]\mathrm{d}\bar{x}
=\frac{L\eta}{F}\int_0^1\left[1+\epsilon^2\lambda^2\left(\frac{\mathrm{d}g}
{\mathrm{d}x}\right)^2\right]\mathrm{d}x,
\end{equation}
and the dimensionless average velocity is given by
\begin{equation}
\begin{split}
U^*&=\frac{\bar{U}^*}{k_BT/(\eta L)}\\
&=\text{Pe}\frac{1}{1+\epsilon^2\lambda^2\int_0^1\left(\frac{\mathrm{d}g}{\mathrm{d}x}\right)^2
\mathrm{d}x}\\
&=\text{Pe}\left\{1-\epsilon^2\lambda^2\int_0^1\left(\frac{\mathrm{d}g}{\mathrm{d}x}\right)^2
\mathrm{d}x + \left[\epsilon^2\lambda^2\int_0^1\left(\frac{\mathrm{d}g}{\mathrm{d}x}\right)^2
\mathrm{d}x\right]^2+O(\epsilon^6)\right\}.
\label{eqn:U*_arclength}
\end{split}
\end{equation}
This simple physical argument recovers the exact results up to $O(\epsilon^2)$. Even for the result at $O(\epsilon^4)$, the effect due to the first derivative of the centerline function is the same.

\section{Conclusion}
We have used the method of asymptotic expansions to calculate two important Macro transport properties in the motion of suspended particles
in a narrow, slowly-varying serpentine channel: the average velocity and the effective diffusion coefficient.
We compare the results for two types of channel, solid-channels that confine the particles with solid walls and soft-channels created by
a confining potential energy landscape.
Our results show that the influence of the solid-channel on the average velocity is the same to that of the soft-channel up to $O(\epsilon^2)$.
Then, the higher order correction, at $O(\epsilon^4)$, shows that the resistance of the solid-channel to particle transport is larger.
However, the difference can be eliminated by changing the width of one of the channels.
Interestingly, in both types of channels, the mobility up to $O(\epsilon^4)$ is independent of the Peclet number and, as a result,
the effective diffusivity satisfies the Einstein-Smoluchowski relation in both cases.

\section{Acknowledgments}
Wang was partially supported by the SC EPSCoR/IDeA GEAR: CRP program and would also like to thank TAPS Fund (Teaching and Productivity Scholarship) and Scholarly Course Reallocation Program at University of South Carolina Upstate for support. Drazer was partially supported by the National Science Foundation Grant No. CBET-1339087.

\end{document}